\documentclass[nofootinbib,amsmath,amssymb,aps,prl,reprint,noeprint,floatfix,superscriptaddress,longbibliography]{revtex4-2}

\usepackage{float}
\usepackage{graphicx}
\usepackage{dcolumn}
\usepackage{soul} 

\usepackage{bm}
\usepackage[version=4]{mhchem}
\usepackage[breaklinks,colorlinks,linkcolor={blue}, %
            citecolor={blue},urlcolor={blue}]{hyperref}
\usepackage{siunitx}
\usepackage{physics}  
\usepackage{amssymb}
\usepackage{amsfonts}                  
\usepackage{graphicx}         
\usepackage[T1]{fontenc}     
\usepackage{ae}                     
\usepackage{lastpage}         
\usepackage{textcomp}
\usepackage{tabularx}
\usepackage{xcolor}
\usepackage{color}
\usepackage{nicefrac}
\usepackage{upgreek}
\usepackage{float}
\usepackage{mathtools}
\usepackage{gensymb}
\usepackage{pdfcomment}

\begin{document}

\title{Excitonic contributions to dark matter-electron scattering}

\author{Nora Taufertsh{\"o}fer}
\affiliation{Materials Theory, ETH Z{\"u}rich, 8093 Zurich, Switzerland}
\author{Vanessa Zema}
\affiliation{Max Planck Institute for Physics, 85748 Garching, Germany}
\affiliation{Institut f{\"u}r Hochenergiephysik der {\"O}sterreichischen Akademie der Wissenschaften, 1010 Wien, Austria}
\author{Riccardo Catena}
\affiliation{Chalmers University of Technology, Department of Physics, SE-412 96 G{\"o}teborg, Sweden}
\author{Valerio Olevano}
\affiliation{CNRS \& Université de Grenoble Alpes, Institut Néel, 38042 Grenoble, France}
\author{Nicola A. Spaldin}
\affiliation{Materials Theory, ETH Z{\"u}rich, 8093 Zurich, Switzerland}

\date{\today}

\begin{abstract}

We determine whether excitonic effects affect predictions of dark
matter (DM)--electron scattering rates by calculating the energy- and momentum-dependent energy-loss function, including electron-hole interaction excitonic effects, for the
dark-matter scintillating detector materials GaAs and NaI. By comparing our results using the
Bethe-Salpeter equation in the framework of many-body perturbation
theory, which explicitly includes excitonic effects, with those
using the quasiparticle random-phase approximation, which includes
only electron-electron interaction and crystal local-field effects, we
find that excitonic effects in NaI significantly increase the predicted
scattering rate at low energy and as a result improve the cross-section sensitivity considering a realistic background.  In contrast, the predicted scattering rate
and the DM-electron scattering cross-section for GaAs are minimally affected
by excitonic effects.

\end{abstract}

\maketitle

Despite decades of direct detection efforts~\cite{Billard_2022_review, Essig_2024_review}, the dark matter (DM) particle has yet to be observed. Until recently, experiments have primarily focused on searching for DM-induced nuclear recoils in low-background detectors, a strategy sensitive only to DM masses above about 1 GeV due to kinematic constraints. The absence of detection events in this mass range has prompted growing interest in sub-GeV DM, which would evade detection via nuclear recoil but could be detectable by its scattering with electrons~\cite{Kopp_DMe:2009, Dedes_DMe:2010, Essig_subGeV_DM:2012}.
Semiconducting targets, for example, in which the scattering process induces electron-hole excitation across the $\sim$ eV band gap, allow sensitivity to DM particles with masses down to $\mathcal{O}$(100 keV).\\
The first numerical predictions of DM--electron scattering rates in semiconducting crystals were made using electronic initial and final states derived from density functional theory (DFT) for silicon and germanium~\cite{Essig_DM_e_semicond:2016}. Many studies followed comparing different target materials~\cite{Griffin_target_comp:2020, Griffin_SiC_DM:2021} and improving or extending the calculations~\cite{Griffin_extended_calc:2021, Chen_DM_Soc:2022, Dreyer_sys_uncertainties:2024}, all within the DFT formalism. Importantly, it was established that the DM problem is separable from the calculation of the electronic structure~\cite{Catena_Emken_DMe:2020, Catena_Emken_DMe_crystal:2021}, and that, for leading DM models that couple the DM to the electron density, the DM-electron scattering rate can be written in terms of the dielectric function of the material and its related energy-loss function (ELF)~\cite{Knapen_dielec:2021, Hochberg_dielec:2021}. These observations mean that the sophisticated machinery developed in the context of theoretical spectroscopy, such as the treatment of in-medium screening effects, collective excitations, and excitonic contributions using first-principles many-body perturbation theory (MBPT) beyond DFT, can be exploited in the calculation of DM-electron scattering rates. Previous works have considered the random phase approximation (RPA)~\cite{Lindhard} on top of the Kohn-Sham electronic structure of DFT with or without crystal local field effects (LFE)~\cite{Knapen_dielec:2021, Knapen_darkelf_2022}, or the $GW$ approximation~\cite{Peterson:2023}, \mbox{which both neglect excitonic effects.}

\begin{figure}[t]
\setlength\belowcaptionskip{-25pt}
\begin{center}
\includegraphics[trim = 0mm 110mm 0mm 6mm, clip, width=1\linewidth]{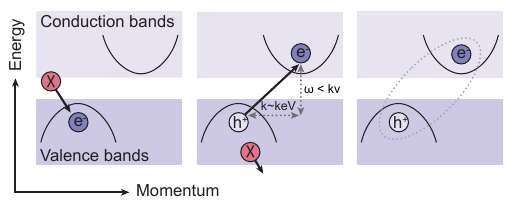}
\caption[]{Schematic of finite-momentum exciton creation by \mbox{DM -- electron} scattering in solid-state targets.}
\label{fig_finite_k_excitons}
\end{center}
\end{figure}
The proposal for direct detection of DM using scintillating materials~\cite{Derenzo_Essig_scint:2017} such as GaAs and NaI has gained new relevance 
since cryogenic scintillating calorimeters have recently reached the required sensitivity and the detection scheme is advantageous for background discrimination~\cite{Zema_light_detec:2024}. The detection signal consists of photons arising from electron-hole recombination following excitation by a DM particle (Fig.~\ref{fig_finite_k_excitons}).
In this mechanism, the attractive interaction between the excited conduction band (CB) electron and the remaining valence band (VB) hole, 
which form an electron-hole bound state called exciton, will likely affect the scattering rate. For example, an increase in the density of states near the CB edge owing to exciton formation might increase the differential DM-electron scattering rate at low deposited energy, particularly for large band-gap scintillators such as NaI. Notably, the usual momentum transfer in a scattering event of an electron and a massive DM particle is of the order of keV~\cite{Derenzo_Essig_scint:2017} $\simeq$ \AA$^{-1}$, far from the optical limit $k \to 0$ studied (with rare exceptions~\cite{Finite_q_excitons_Gatti:2013}) in the context of excitons in solid-state physics.
To date, these excitonic effects have not been examined in the DM literature. 
Since NaI and GaAs are the detector materials used by experiments that are specifically designed to measure scintillation signals (NaI in COSINUS ~\cite{cosinus_exp:2024, Angloher_cosinus:2025}; GaAs in TESSERACT~\cite{snowmass_tesseract:2021} and DAREDEVIL~\cite{Helis_daredevil_GaAs:2024}), determining the impact of electron-hole interactions in these materials is necessary for reliable theoretical support.

Motivated by these considerations, here we solve the Bethe-Salpeter equation (BSE)~\cite{SalpeterBethe51, HankeSham79, Strinati_BSE:1988, OnidaAndreoni95} to calculate the frequency and momentum dependent dielectric function of GaAs and NaI within \textit{ab initio} MBPT and describe their target response in DM-electron scattering events. With respect to previous works limited to the RPA level, the BSE inherently captures electron-hole interaction excitonic effects.
By comparing our BSE and RPA calculations, we are able to quantify the impact of finite-momentum excitonic effects on direct DM searches with NaI and GaAs targets. 
Our main finding is that for both materials, the BSE approach predicts a larger ELF at low energies compared to that of RPA across a large finite-momentum range, and that the effect is much more pronounced in NaI. 
This property of the BSE ELF increases the DM-electron scattering rate at energies of the order of the band gap. The increase in the BSE rate in turn leads to an almost one order of magnitude higher predicted cross-section sensitivity in NaI for DM particles with a mass larger than 2 MeV and including a realistic background rate. Our findings suggest that for large band gap scintillators the inclusion of excitonic effects is essential for predicting the DM-electron scattering rate at low energy and the cross-section sensitivity over the light DM mass range. 
Conversely, from a condensed matter point of view, scattering with DM will allow excitation of finite-momentum excitons that are unavailable (dark) to optical absorption and, for the highest momenta, outside the range of X-ray scattering spectroscopy.

This paper proceeds as follows: First, for the benefit of the condensed-matter community, we summarize the formalism for calculating the DM-electron scattering rate in terms of the dielectric function, following Ref.~\cite{Knapen_dielec:2021}. 
Second, for the benefit of the DM community, we briefly introduce the MBPT formalism and the approaches used to calculate the dielectric function, namely BSE and RPA with or without LFE.
Next, we show our results for the ELF and the differential scattering rates calculated with both methods for the two target materials NaI and GaAs. 
Finally, we present the predicted DM-electron scattering cross-section sensitivity over the mass range of light DM and discuss the implications of our results for the prospects of a dedicated experiment.

As stated above, for the case of non-relativistic DM, $\chi$, coupling to the electron density in a target material, the DM-electron scattering rate can be formulated in terms of the longitudinal frequency- and momentum-dependent dielectric function $\epsilon(\omega, \mathbf{k})$~\cite{Hochberg_dielec:2021, Knapen_dielec:2021}. 
\footnote{In principle we can also consider the coupling with the electron current for the case, e.g.,  of a vector boson mediator. This case requires the calculation of the full dielectric tensor, including transverse components.}
An additional field $\phi$, which can be a massive or massless, scalar or vector boson, e.g. a dark photon, mediates the (weak) coupling between the standard-model electron and the DM $\chi$.
The rate is then given by~\cite{Knapen_dielec:2021}
\begin{align}
\label{eq:rate_from_ELF}
   R &= \frac{1}{\rho_T}
   \frac{\rho_{\chi}}{m_\chi} \frac{\pi}{\alpha_{\text{em}}}
    \frac{\overline{\sigma}_e}{\mu_{\chi e}^2 }
     \int d^3 v f_\chi(v) \nonumber  
    \times \int \frac{d^3 k}{(2\pi)^3} k^2 |F_{\phi}(k)|^2 \\
    & \int \frac{d\omega}{2\pi} \frac{1}{1 - e^{-\beta \omega}} \text{Im} \left[ \frac{-1}{\epsilon(\omega, \mathbf{k})} \right] \delta \left( \omega + \frac{k^2}{2 m_\chi} - \mathbf{k} \cdot \mathbf{v} \right) \, ,
\end{align}
where $\rho_{\text{T}}$ is the target density, $\rho_\chi=0.4$~GeV/cm$^3$ the local DM density, $m_{\chi}$ the variable DM mass, $\alpha_{\text{em}}$ the fine structure constant, $\beta$ the inverse temperature, $\mu_{\chi e}$ the reduced mass for DM and electron and $\overline{\sigma}_e$ a reference cross-section defined in the Supplemental Material (SM). 
In this work, we consider two limiting cases for the DM-mediator form factor, namely $F_{\phi}(k) = \alpha^2_{\text{em}} m_e^2/k^2$ for a massless mediator where $m_e$ is the electron mass and $F_{\phi}(k) = 1$ for a massive mediator.
The rate includes an integration over the DM velocity distribution $f_{\chi}(v)$, which is assumed to follow the standard halo model with details given in the SM. Finally, the term in Eq.~(\ref{eq:rate_from_ELF}) containing the inverse dielectric function $-\Im \varepsilon^{-1}(\omega, \mathbf{k})$ is precisely the ELF, which captures the rate at which an electron loses energy and momentum when traversing a target dielectric material.
We emphasize again that this formulation for the DM-electron scattering rate distills the electronic part of the problem into one quantity, the dielectric function $\epsilon(\omega, \mathbf{k})$, which can be calculated using condensed-matter \textit{ab initio} theories. The level of theory can then be selected depending on which physical effects, such as in-medium screening, or single-particle or collective excitations are important for determining the DM-electron rate.
The relevant energy and momentum range can also be selected, noting that the Dirac delta in Eq.~(\ref{eq:rate_from_ELF}) ensures the kinematic relation for the energy deposition as a function of the transferred momentum $\mathbf{k}$, $\omega = \mathbf{k} \cdot \mathbf{v} - k^2/(2m_{\chi})$~\cite{Essig_DM_e_semicond:2016}.
Hence, the energy deposited in a DM-electron scattering event is constrained by $\omega < k\, v$.
Within the employed standard halo model, the DM velocity is bounded by \mbox{$v \lesssim v_{\text{esc}}+v_{\text{Earth}} =740\, \text{km/s}$}~\cite{Derenzo_Essig_scint:2017}. For example, a transferred momentum of $k=10$~keV therefore restricts the energy deposition to $\omega_{\text{k}=10\,\text{keV}}\leq25$\,eV.
This energy range, in which excitonic effects are typically important in semiconducting systems, further justifies our choice of the BSE to calculate the ELF.\\
In a compact form, the BSE can be written~\cite{Strinati_BSE:1988}
\begin{align}
\label{eq:BSE}
  L = L^0 + L^0 \Xi L
,
\end{align}
where $L(x_1,x_2;x_3,x_4)$ is the two-particle correlation function (with $x_i = (t_i,\mathbf{r}_i)$ the space-time four-vector), $L^0(x_1,x_2;x_3,x_4) = G(x_2,x_4)G(x_3,x_1)$ with $G$ the one-particle Green function, and the BSE kernel $\Xi = \delta\Sigma/\delta G$ defined as the functional derivative of the irreducible self-energy $\Sigma$ with respect to $G$.
In the $GW$ approximation for the mass-operator, $\Sigma_M = i G W$, the kernel becomes $\Xi(x_1,x_2;x_3,x_4) = -i \delta(x_1,x_2) \delta(x_3,x_4) \delta(t_1,t_3) v(\mathbf{r}_1,\mathbf{r}_3) +i \delta(x_1,x_3)\delta(x_2,x_4)\delta(t_1,t_2) W(\mathbf{r}_2,\mathbf{r}_1,\omega=0)$ in terms of the bare Coulomb interaction $v$ and the statically ($\omega=0$) screened electron-hole attraction $W$.
In the standard approach employed in this work, the screening $W^\mathrm{RPA}$ is derived from a separate RPA calculation. 
Here, we solve the BSE within the Tamm-Dancoff approximation (TDA), which neglects the coupling between the resonant and the anti-resonant parts, i.e. it neglects electron-hole pairs propagating backward in time.
Once the BSE is solved for $L$, we calculate the polarizability $\chi(x_1,x_2) = -i L(x_1,x_2;x_1^+,x_2^+)$ (where $x_i^+ = (t_i + 0^+, \mathbf{r}_i)$), and the inverse microscopic dielectric function $\varepsilon^{-1} = 1 + v\chi$. The Fourier transform  then provides the macroscopic dielectric function including LFE $\epsilon_M(\mathbf{k}=\mathbf{q}+\mathbf{G},\omega) = 1/\varepsilon^{-1}_\mathbf{GG}(\mathbf{q},\omega)$ (with $\mathbf{q}$ within the first Brillouin zone and $\mathbf{G}$ a reciprocal lattice vector) which enters Eq.~(\ref{eq:rate_from_ELF}).
We compare the BSE results capturing excitonic effects to RPA which is retrieved by neglecting the electron-hole screened interaction $W$ in the kernel $\Xi$ of Eq.~(\ref{eq:BSE}).
Excluding crystal LFE in RPA (or in BSE) neglects microscopic contributions of the induced field to the macroscopic response, so that an independent (quasi)particle picture is retrieved.\\
To perform the electronic structure calculations, we use the full-potential all-electron code \texttt{exciting}~\cite{Gulans_exciting:2014, Vorwerk_exciting_bse:2019} which allows us to avoid the difficulties associated with the widely-used pseudo-potential approximation~\cite{Griffin_extended_calc:2021} and to directly capture the high-momentum components of the valence and semicore states.
First, we carry out ground-state DFT calculations within the local density approximation (LDA) for both GaAs and NaI with details given in the SM.
We then apply quasiparticle corrections to the DFT Kohn-Sham electronic structure.
For coherence with the BSE kernel, this is usually done by a $GW$ self-energy. Here, for the sake of comparison with previous literature, we used a scissor operator matched to reproduce the experimental band gaps $E_g^\mathrm{GaAs} = \SI{1.519}{eV}$~\cite{Grilli_GaAs_band_gap} and $E_g^\mathrm{NaI} = \SI{5.9}{eV}$~\cite{NaI_band_gap_Brown:1970}.
The resulting band structures are shown in the SM. \\
For a valid comparison of the ELFs from RPA and BSE, we use the same underlying scissor-corrected Kohn-Sham states as starting point for both methods. 
As the BSE approach is computationally expensive, scaling to the third power of the number of k-points and included bands, we opted for a balanced setup in terms of VBs, CBs, and k-grid resolution.
A 6x6x6 k-grid and up to 14 bands in total offer a good compromise between accuracy and cost. For GaAs (NaI), selecting 4~VBs (3~VBs) and 10~CBs allows us to capture all electron-hole excitations in an energy range spanned by the highest VB and CB up to $\sim$12\,eV. 
We then use the same computational parameters tested for BSE in RPA, with the exception that we include the semicore states which, however, do not contribute in the low-energy range of interest. In the SM we briefly discuss excitonic effects for the Ga 3$d$ semicore states when a larger energy range is considered.\\
For deriving the DM-electron scattering rates we use the open-source code \texttt{DarkELF}~\cite{Knapen_darkelf_2022, DarkElf_code} with our computed ELFs and restrict our study to isotropic responses, in which the dielectric function depends only on the absolute value of the momentum $|\mathbf{k}|$. 

\begin{figure*}[htbp]
\begin{center}
\includegraphics[trim = 0mm 95mm 0mm 0mm, clip, width=1\linewidth]{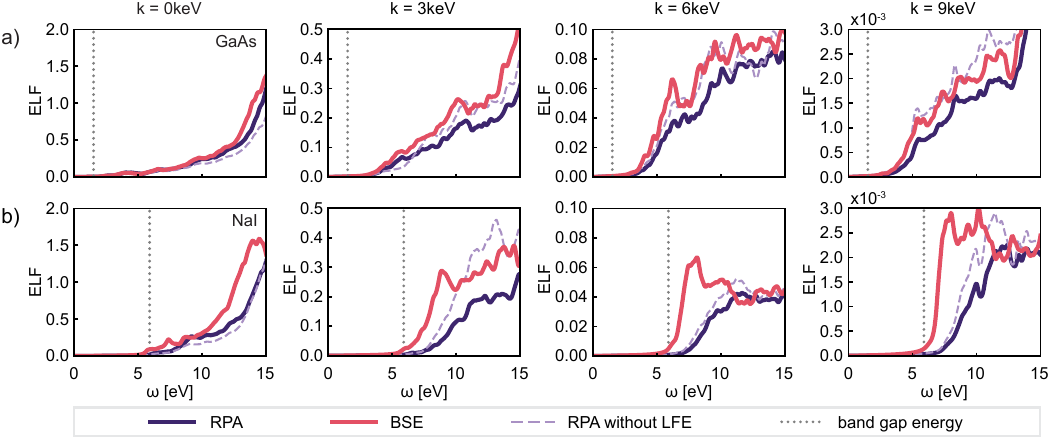}
\end{center}
\caption{ELF at various transferred momenta calculated with RPA and BSE for a) GaAs and b) NaI. At low energies, the ELF from BSE is always larger than that from RPA due to the excitonic contributions, with the largest difference for NaI. The ELF for RPA without LFE corresponds to independent particle transitions. }
\label{fig_ELFs_GaAs_NaI_various_k_same_settings}
\end{figure*}
\begin{figure}[htbp]
\setlength\belowcaptionskip{-10pt}
\begin{center}
\includegraphics[trim = 0mm 78mm 0mm 0mm, clip, width=1\linewidth]{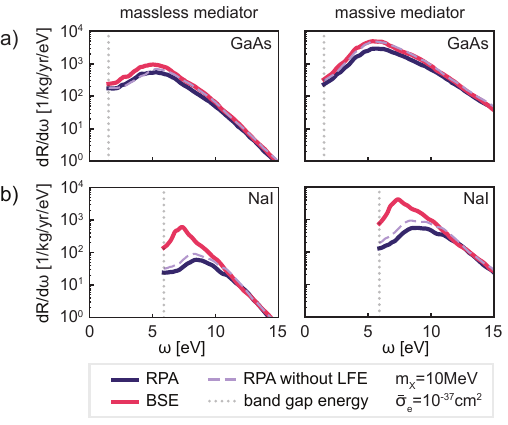}
\caption[]{Differential scattering rates for a) GaAs  and b) NaI calculated with RPA and BSE for the limiting cases of a massles and a massive mediator. The underlying dielectric functions include 10~CBs and transferred momenta $|\mathbf{k}|= [\text{0, 0.3, 0.7, 1.0,..., 14.7}]$\,keV.}
\label{fig_diff_rates}
\end{center}
\end{figure}
\begin{figure}[t]
\begin{center}
\includegraphics[trim = 0mm 75mm 0mm 0mm, clip, width=1\linewidth]{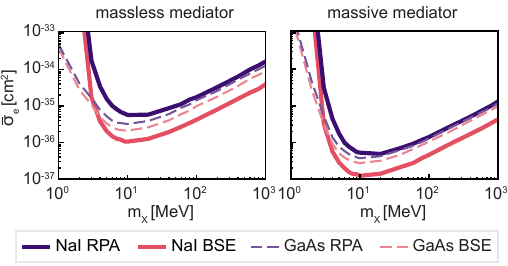}
\caption[]{Reach over the light dark matter range for GaAs and NaI for a kg-year exposure assuming a monotonically decaying background. The BSE approach predicts a better sensitivity, in particular for NaI.}
\label{fig_reach}
\end{center}
\end{figure}

We now compare our ELFs calculated within RPA and BSE for the two materials GaAs and NaI, using 10~CBs in each case for consistency. Fig.~\ref{fig_ELFs_GaAs_NaI_various_k_same_settings} shows the ELF in the low-energy range up to $\omega=\SI{15}{eV}$ in the optical limit ($k=0$) and at three non-zero values of momentum transfer ($k=3, 6 \text{ and } \SI{9}{\kilo\electronvolt}$). For both materials, the BSE ELF is larger than the RPA ELF in this energy range due to the excitonic contributions. The BSE increase is strong for NaI, particularly at energies near the band gap, while in GaAs it is smaller and of similar size over the displayed energy range. 
For NaI, the lowest excitonic contributions to the ELF below the band gap energy are clearly visible for $k=0$ and $k=\SI{3}{\kilo\electronvolt}$.
Excluding LFE in RPA leads to an increased ELF at finite momentum that even partially exceeds the BSE ELF at larger energies. For technical details, we refer to the SM.\\
The properties of the ELF are directly transferred to the scattering rates which we present next. 
Fig.~\ref{fig_diff_rates} shows the differential scattering rates for the massless and massive mediator case including transferred momenta up to $\SI{14.7}{\kilo\electronvolt}$. The DM mass and reference cross-section are fixed to $m_{\chi}=\SI{10}{\mega\electronvolt}$ and $\overline{\sigma}_e = \SI{1e-37}{\centi\meter\squared}$, respectively.
For GaAs, the BSE rate is only slightly larger at low energies compared to that of RPA and the overall shape is similar as expected from Fig.~\ref{fig_ELFs_GaAs_NaI_various_k_same_settings} a).
In contrast, for NaI, we observe a significantly larger BSE rate which peaks at about $\SI{8}{\electronvolt}$. This feature is due to the large contributions of the ELF at higher momentum transfers, as shown in Fig.~\ref{fig_ELFs_GaAs_NaI_various_k_same_settings} b).
Excluding LFE in RPA slightly increases the rate since local screening effects are neglected. For GaAs and a massive mediator, the BSE rate and that for RPA without LFE even overlap due to the similar ELF at finite momentum.\\
Next, we present the cross-section sensitivity for GaAs and NaI based on our rate calculations using BSE and RPA over the light DM mass range. For a kg-year exposure we assume a monotonically decaying experimental background in the light detector~\cite{cresst_background:2023} with the number of events of the order $10^9$ including a suppression factor of $\lambda_{\text{bck}}=10^{-4}$, following the method in Zema et al.~\cite{Zema_light_detec:2024}. Further details are given in the SM.
For NaI, we assume an absorption onset at $\SI{5.6}{\electronvolt}$, corresponding to the lowest energy exciton absorption peak in optical measurements~\cite{NaI_exciton_absorption_Teegarden:1967}, and $\SI{4.2}{\electronvolt}$ for emission, corresponding to electron-hole recombination from the self-trapped exciton~\cite{Selftrapped_ex_Williams:1990}. For GaAs we use the measured band gap value of \SI{1.519}{eV}~\cite{Grilli_GaAs_band_gap} for absorption, justified by the small binding energy of $\SI{\sim3}{\milli\electronvolt}$ of the lowest optical exciton~\cite{GaAs_exciton_binding_Sturge:1962}, and the reported value $\SI{1.48}{\electronvolt}$ for emission~\cite{Vasiukov_GaAs_scintill:2019}.
We consider a combined light yield and collection efficiency of $\epsilon_l =13\%$ according to Ref.~\cite{COSINUS:2017bco} for NaI and choose the same value for GaAs for comparison. Note that doped GaAs can achieve light yields that are significantly higher than those of pure crystals~\cite{Vasiukov_GaAs_scintill:2019}. 
We plot the resulting reach for both massless and massive mediators over a range of light DM masses in Fig.~\ref{fig_reach}, and see that the BSE method yields a better sensitivity than the RPA for the full mass range. While this improvement is moderate for GaAs, excitonic effects in NaI improve the reach by almost an order of magnitude. 
The smallest considered DM mass cannot be probed with NaI as the low-energy threshold in the scattering
rate is still almost 4 times larger than in GaAs, even when accounting for the lowest exciton.

In this work, we showed that excitonic effects captured in the BSE ELF increase the predicted DM-electron scattering rate at low energy compared with predictions from the RPA method. This effect is considerably more pronounced for NaI, which has an exciton binding energy that is three orders of magnitude larger than that of GaAs in the optical limit. Therefore, our results suggest that electron-hole interactions in scattering events with finite momentum transfer have a greater impact in materials with stronger bound optical excitons. Importantly, we found that excitonic effects are significant even at large momentum transfer, where intuition from optical experiments is limited.
The discovery of an increased BSE DM-electron scattering rate allowed us to report improved cross-section sensitivities for both GaAs and NaI detectors considering a realistic background rate. 
The substantial improvement of the reach for NaI highlights that accounting for excitonic effects is necessary for accurately describing the response in this target material.
We note that neither coupling to antiresonant electron-hole transitions beyond the TDA at finite momentum transfer, nor correlation effects beyond the $GW$ approximation to the BSE adiabatic-only kernel $\Xi$, are included in our calculations, opening a perspective for future theoretical developments. Likewise, non-radiative relaxation processes that lead to the formation of self-trapped excitons in NaI~\cite{Selftrapped_ex_Williams:1990, Nagata_excitons_NaI:1990} and the energy dependence of the emission are not studied here. While self-trapping of excitons could in principle be captured in first-principles calculations~\cite{Dai_ab_initio_STE}, both processes might be more effectively incorporated through phenomenological modeling based on experimental measurements.
Indeed, the measurement of electronic excitations in cryogenic scintillating calorimeters is the scope of a dedicated detector and phenomenology development  -- the O$\nu$DES project -- in collaboration with the COSINUS experiment~\cite{Zema_light_detec:2024, Angloher_cosinus:2025}. Our work emphasizes the importance of such accurate calibration of the cryogenic scintillating calorimeters for DM-electron scattering searches, as well as the desirability of calibration sources inducing finite-momentum excitons over light sources. \\

We thank Tanner Trickle, Karoline Schäffner and Paolo Settembri for helpful discussions. NT, RC, and NAS acknowledge support from the Swiss National Science Foundation under grant no. 200021-231539. NAS and VO thank the CNRS Fellow Ambassador program for supporting NAS's visit to the Institut Néel. VZ acknowledges the support of the Klaus Tschira foundation to the O$\nu$DES project. Computational resources were provided by the ETH Zurich Euler cluster.
The relevant input files of our ab initio calculations and the data supporting the findings of this work will be made publicly available upon publication of the manuscript. Until then they are available upon request. \vfill\eject

\bibliography{references}

\end{document}


\title{Supplemental Material: Excitonic contributions to dark matter-electron scattering}

\author{Nora Taufertsh{\"o}fer}
\affiliation{Materials Theory, ETH Z{\"u}rich, 8093 Zurich, Switzerland}
\author{Vanessa Zema}
\affiliation{Max Planck Institute for Physics, 85748 Garching, Germany}
\affiliation{Institut f{\"u}r Hochenergiephysik der {\"O}sterreichischen Akademie der Wissenschaften, 1010 Wien, Austria}
\author{Riccardo Catena}
\affiliation{Chalmers University of Technology, Department of Physics, SE-412 96 G{\"o}teborg, Sweden}
\author{Valerio Olevano}
\affiliation{CNRS \& Université de Grenoble Alpes, Institut Néel, 38042 Grenoble, France}
\author{Nicola A. Spaldin}
\affiliation{Materials Theory, ETH Z{\"u}rich, 8093 Zurich, Switzerland}

\date{\today}

\maketitle

\setcounter{section}{0}
\setcounter{equation}{0}
\setcounter{figure}{0}
\setcounter{table}{0}
\setcounter{page}{1}
\makeatletter

\renewcommand{\theequation}{S\arabic{equation}}
\renewcommand{\thefigure}{S\arabic{figure}}
\renewcommand{\thetable}{S-\Roman{table}}
\renewcommand{\bibnumfmt}[1]{[S#1]}
\renewcommand{\citenumfont}[1]{S#1}

\section{Ground-state properties}

\begin{figure*}[t!]
\begin{center}
\includegraphics[trim = 10mm 185mm 0mm 15mm, clip, width=1\linewidth]{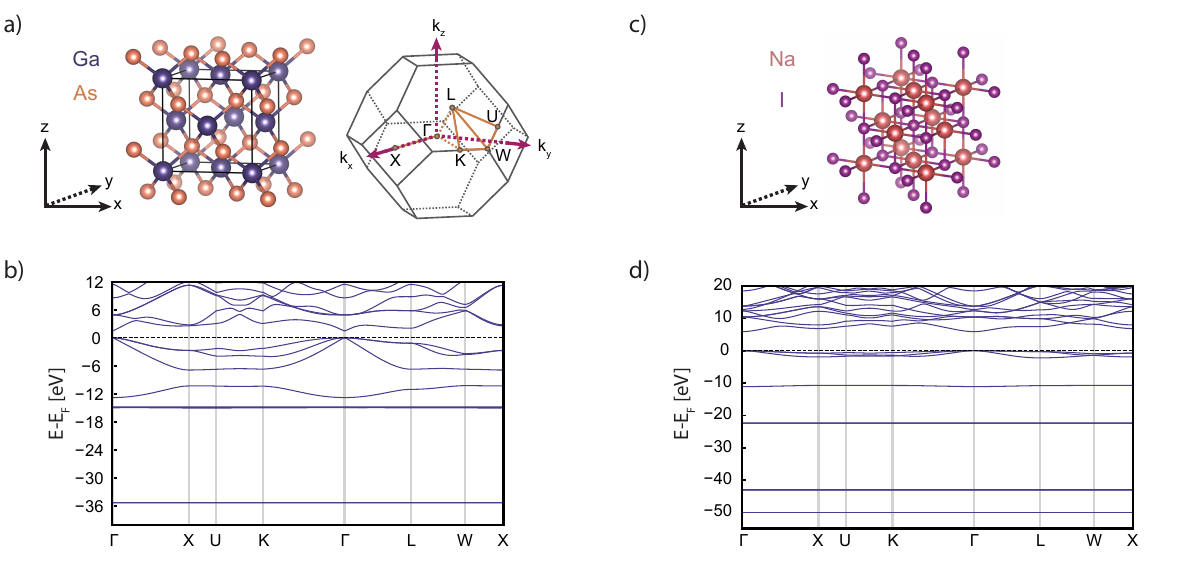}
\caption[]{Crystal structures visualized with \texttt{vesta}~\cite{VESTA_Momma:2008}, first Brillouin zone (BZ) and band structures for GaAs and NaI, obtained from DFT LDA by a scissor-operator correction to match the respective experimental band gap. 
a) Zinc blende crystal structure of GaAs and first BZ with labeled high symmetry points. b) GaAs band structure including the Ga (upper) and As (lower) 3$d$ semicore states. c) Rock salt crystal structure of NaI. d) NaI band structure including all semicore states: from bottom upward Na 2$s$, I 4$d$, Na 2$p$, and I 5$s$. BZ and high symmetry paths coincide with those of GaAs. }
\label{fig_crystal_band_struct}
\end{center}
\end{figure*}

We perform ground-state density functional theory (DFT) calculations within the local density approximation (LDA) for both GaAs and NaI with experimental lattice constants $a_{\text{GaAs}}=\SI{5.653}{\angstrom}$~\cite{Zhou_GaAs_lattice:2001} 
and \mbox{$a_{\text{NaI}}=\SI{6.475}{\angstrom}$ ~\cite{handbook_optics_NaI_lattice:2010}} for the cubic cells that are shown in Fig.~\ref{fig_crystal_band_struct} a) and c). In our calculations we use primitive cells with a 6x6x6 k-grid and $\text{rgkmax}=9$. The latter parameter implicitly sets the basis size in \texttt{exciting}, as it represents the product $R_{\text{min}} \cdot |\mathbf{G}+\mathbf{k}|_{\text{max}}$ with $R_{\text{min}}$ the smallest of all muffin-tin radii.
We include all semicore states, i.e. Ga 3$d$ and As 3$d$ states for GaAs and Na 2$s$, 2$p$ and I 4$d$, 5$s$ for NaI, in our self-consistent calculation. The corresponding scissor-corrected band structures are displayed in Fig.~\ref{fig_crystal_band_struct} b) and d).
%

\section{Details on the ELF calculation}

\begin{figure}[t]
\begin{center}
\includegraphics[trim = 0mm 72mm 0mm 0mm, clip, width=1\linewidth]{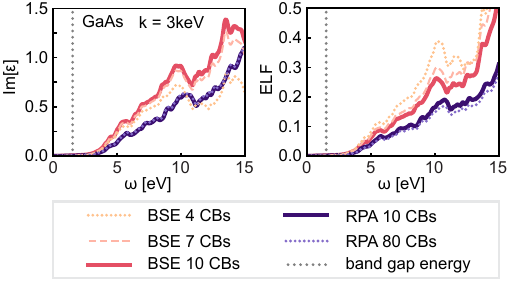}
\caption[]{$\text{Im}(\epsilon)$ and ELF of GaAs for a range of considered CBs to illustrate the convergence behavior of the absorption and the ELF with respect to the number of CBs. }
\label{fig_eps_ELF_convergence_CB_RPA_BSE}
\end{center}
\end{figure}

We calculate the dielectric function for transfer momenta along the (100) direction with $|\mathbf{k}|=[0, 0.3, 0.7, 1.0,...,14.7]$\,keV and subsequently derive the energy loss function (ELF). For the input in \texttt{DarkELF}, the ELF is then approximated to be isotropic.
For NaI, we also compute within the random-phase approximation (RPA) the dielectric function with momentum transfer along the inequivalent (111) direction. Although we observed a slight change in the oscillator strength of the ELF we confirmed that this change is too small to be reflected in the differential scattering rate.\\
In order to recognize excitonic effects when comparing the ELF from the Bethe-Salpeter equation (BSE) and RPA and to minimize differences only due to numerical settings, we choose equal parameters in both methods. 
We note that for GaAs a denser k-grid than 6x6x6 smoothens the calculated ELF but does not change the general shape or magnitude. For NaI, a denser k-grid does not lead to an observable change in the ELF as the bands are less dispersive than in GaAs.\\
In RPA, we are significantly less restricted by computational cost, which allows us to include all semicore states and additional conduction bands (CBs). Including more CBs extends the range of the absorption spectrum to higher energies but leaves the lower energy region, where transitions to higher CBs do not yet contribute, unaffected. 
It also shifts the plasmons (the zeros of the real part of the dielectric function, related to the imaginary part via the Kramers-Kronig relation) to higher energies, such that the ELF is decreased at low energy. This effect becomes apparent in  Fig.~\ref{fig_eps_ELF_convergence_CB_RPA_BSE} when we compare the RPA absorption and the ELF with 10~CBs to that with 80~CBs for a momentum transfer of \SI{3}{\kilo\electronvolt} in GaAs. The same properties apply to BSE, for which we can only check the convergence between 4~CBs and up to 10~CBs.\\
In this work, we solve the BSE within the Tamm-Dancoff approximation (TDA), which neglects the coupling between the resonant and the anti-resonant parts. A BSE calculation in the optical limit beyond TDA revealed that for GaAs there is no sizeable change in the ELF. For NaI, we see an increase in the ELF compared to TDA. We leave the incorporation of finite momentum excitonic effects from BSE beyond TDA for future work.\\
%
%
\section{Details on the dark matter model}

In the following, we provide more details on the dark matter (DM)-electron scattering rate equation stated in the main text.
The DM-mediator form factor is defined by
\begin{align}
    F_{\phi}(k) = \frac{\alpha_{\text{em}}^2 m_e^2 + m^2_{\phi}}{k^2 + m^2_{\phi}}
\end{align}
for which we consider the two limiting cases $m_{\phi} \rightarrow 0$ for a massless and $m_{\phi} \rightarrow \infty$ for a massive mediator.
The effective cross-section is given by
\begin{align}
    \bar{\sigma}_e = \frac{\mu^2_{\chi e} g^2_e g^2_{\chi}}{\pi(\alpha_{\text{em}}^2 m_e^2 + m^2_{\phi})^2} \,
    \label{eq:refcross}
\end{align} 
where $g_e$ ($g_{\chi}$) defines the coupling strength between the mediator and the electron \mbox{number density (DM) \cite{Knapen_dielec:2021}.}
The DM velocity distribution $f_{\chi}(v)$ is assumed to follow the standard halo model with the most probable speed $v_0=220$~km/s, 
the galactic escape velocity $v_{\text{esc}}=500$~km/s defining the minimum velocity for DM to escape the gravitational potential well of the halo, and the average velocity of Earth in the galactic rest frame $v_{\text{Earth}}=240$~km/s. We choose these parameter values of the velocity distribution for comparison with previous literature results.

\section{Contribution of semicore states to the rate }
\label{app:semicore_states_rate}

\begin{figure}[htbp]
\begin{center}
\includegraphics[trim = 0mm 80mm 0mm 0mm, clip, width=1\linewidth]{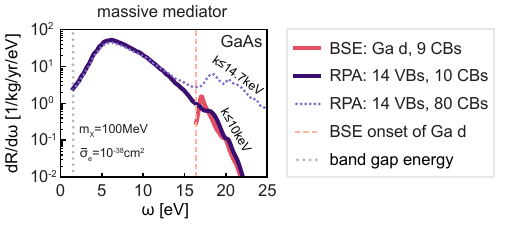}
\caption[]{Differential scattering rate in GaAs for a massive mediator and larger energy range to show the contribution of electronic transitions from the Ga $d$ semicore states.}
\label{Ga_d_excitonic_effect}
\end{center}
\end{figure}

For predictions that include high-energy depositions far from the band-gap energy, the importance of incorporating semicore and core states in the rate calculation  was discussed in \cite{Griffin_extended_calc:2021}. Although in our work, the main focus lies on the impact of excitonic effects expected to be most prominent near the low energy threshold, we briefly address the influence of electron-hole interaction for transitions from semicore states.
In Fig.~\ref{Ga_d_excitonic_effect}, we compare the RPA differential rate for a massive mediator to a BSE calculation considering the Ga $d$ semicore states and 9~CBs, with the momentum transfer restricted to $k\leq$ \SI{10}{\kilo\electronvolt}. In BSE, the onset of the contribution from the Ga 3$d$ states occurs close to $\sim \SI{16.4}{\electronvolt}$ which is the sum of the binding energy of the Ga $d$ states and the band gap energy. In RPA, the rate increases only at slightly larger energy and the increase is significantly less peaked.
By incorporating transfer momenta up to \SI{14.7}{\kilo\electronvolt} and 80~CBs in RPA we loosen the kinematic restriction $\omega < kv$. Therefore, the rate does not drop rapidly at energies beyond \SI{15}{\electronvolt} and reveals better the contributions due to the semicore states. The onset of the semicore contribution at lower energy in BSE is due to the electron-hole interactions that push oscillator strength to lower energies.

%
%
\section{Details on the reach calculation}
\label{app:details_reach}

\begin{figure}[t!]
\begin{center}
\includegraphics[trim = 0mm 75mm 0mm 0mm, clip, width=1\linewidth]{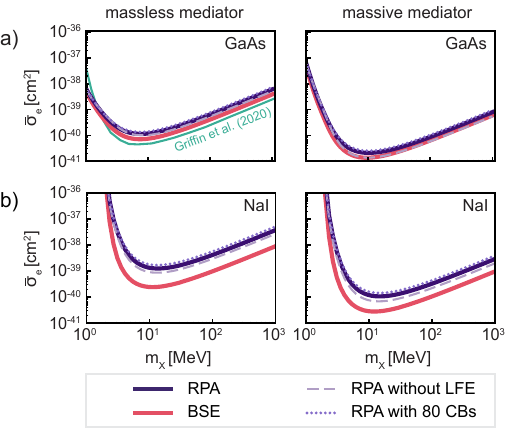}
\caption[]{Cross-section sensitivity over the light dark matter range for a) GaAs and b) NaI assuming three events for a kg-year exposure, zero background and the absorption onset as the sensitivity threshold. For comparison, we also show the sensitivity for GaAs without in-medium screening and a light dark photon mediator in green as reported in \cite{Griffin_target_comp:2020}. }
\label{fig_reach_only_theory}
\end{center}
\end{figure}

\begin{figure}[t!]
\begin{center}
\includegraphics[trim = 0mm 75mm 0mm 0mm, clip, width=1\linewidth]{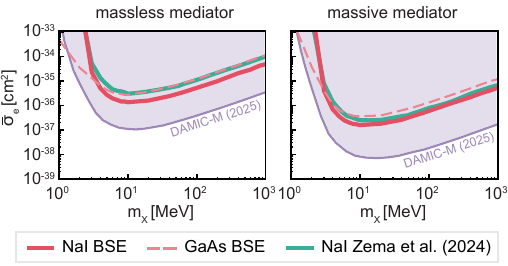}
\caption[]{Reach over the light DM range assuming a kg-year exposure, a monotonically decaying background suppressed by $\lambda_{\text{bck}}=10^{-4}$ and $\epsilon_l =10\%$ for comparison with the sensitivity in NaI reported by Zema et al.~\cite{Zema_light_detec:2024} where screening and excitonic effects were neglected. Recent 90$\%$ C.L. upper limits from DAMIC-M are taken from~\cite{DAMIC-M_exclusion:2025}.}
\label{reach_BSE_vs_QEdark_vs_DAMIC}
\end{center}
\end{figure}

In Fig.~\ref{fig_reach_only_theory} we provide the cross-section sensitivities neglecting any background, which are derived by integrating the differential scattering rates over the deposited energy within \texttt{DarkELF}~\cite{Knapen_darkelf_2022}, assuming three events for a kg-year exposure and the absorption onset at $\SI{5.6}{\electronvolt}$~\cite{NaI_exciton_absorption_Teegarden:1967} ($\SI{1.519}{\electronvolt}$~\cite{Grilli_GaAs_band_gap}) for NaI (GaAs) as low-energy threshold.\\
In addition to the choice of 10~CBs, employed in the main text for comparability between BSE and RPA, we also show the RPA cross-section sensitivity for 80~CBs. The higher number of CBs slightly increases the cross section, which is due to the decrease of the ELF as shown in Fig.~\ref{fig_eps_ELF_convergence_CB_RPA_BSE} and hence the rates. For GaAs, we include the sensitivity reported by Griffin et al.~\cite{Griffin_target_comp:2020} for a light dark photon mediator. The sensitivity is better compared to both our RPA and BSE results as the calculation neglects any in-medium screening. While reduction in sensitivity due to screening effects has been previously reported, e.g. in~\cite{Knapen_dielec:2021}, we note that the excitonic effects captured in BSE counteract this trend.\\ 
\begin{figure}[b!]
\begin{center}
\includegraphics[trim = 0mm 75mm 0mm 0mm, clip, width=1\linewidth]{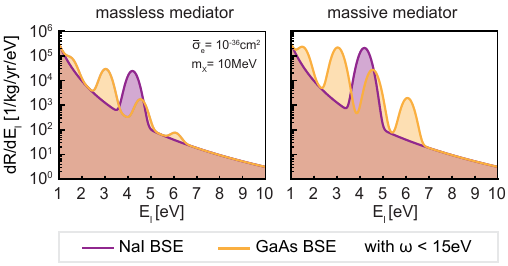}
\caption[]{BSE single photon peaks induced by DM-electron scattering combined with the background rate detected in the light detector with an energy resolution of $\sigma^{\text{LD}}_{\text{res}}=\SI{0.2}{\electronvolt}$. The background is taken from Fig. 4 in~\cite{cresst_background:2023} and suppressed by $\lambda_{\text{bck}}=10^{-8}$ for visualization.}
\label{peak_plot}
\end{center}
\end{figure}
For experiments that measure scintillation light, a more realistic cross-section sensitivity is obtained by considering an experimental background rate in the light detector~\cite{cresst_background:2023}, following the method in~\cite{Zema_light_detec:2024} with a profile-likelihood-based test and Monte Carlo data for the background-only and signal-and-background hypothesis. Any background due to the scintillating target is neglected, as the so-called low-energy excess in the light detector dominates.
The expected energy in the light detector is $E_{l}=n_{\gamma} E_{\text{em}}$ with $n_{\gamma}$ the number of detected photons and $E_{\text{em}}$ the photon emission energy.
The electron energy threshold to generate $n_{\gamma}$ photons is \mbox{$E_{w,n_\gamma} = E_{\text{abs}}+(n_\gamma-1) \langle E \rangle$}, with $E_{\text{abs}}$ the energy threshold for absorption and $\langle E \rangle$ the mean energy to produce two or more scintillation photons, fixed to $\langle E \rangle = 3(2.764)E_{\text{abs}}$ for NaI (GaAs), according to ~\cite{Derenzo_Essig_scint:2017}. 
The signal rate in the light detector is given by
\begin{align}
    \frac{\text{d}R}{\text{d}E_l}
    = \int dE'\  R_{n_\gamma}(\bar{\sigma}_e, m_\chi) \cdot \delta(E'-n_\gamma E_{\text{em}})\cdot f(E_l-E')
    \label{eq:dRdEl}
\end{align} 
with 
\begin{align}
    R_{n_\gamma}(\bar{\sigma}_e, m_\chi) = \int^{E_{\omega, n_{\gamma}+1}}_{E_{\omega, n_\gamma}} \frac{\text{d}R}{\text{d}\omega} (\bar{\sigma}_e, m_\chi) \ \text{d}\omega
    \label{eq:Rngamma}
\end{align} 
and $f$ a Gaussian energy resolution function of width $\sigma^{\text{LD}}_{\text{res}}$.
The light yield and collection efficiency are taken into account by multiplying the rate by a constant factor $\epsilon_l$ in energy, set at the experimental value of 13\%~\cite{COSINUS:2017bco}.
As an example, in Fig.~\ref{peak_plot} we show the rate of scintillation photons according to Eq.~(\ref{eq:dRdEl}) as a function of deposited energy in the light detector based on our BSE differential scattering rates ($\text{d}R/\text{d}\omega$) for GaAs and NaI for a reference cross section $\overline\sigma_e = 10^{-36}$ cm$^2$ and DM mass $m_{\chi}= 10$ MeV. 
The peaks refer to the scintillation photons detected in the light detector with $\sigma^{\text{LD}}_{\text{res}}=\SI{0.2}{\electronvolt}$. This energy resolution is chosen such that the threshold for observing a peak, which is assumed to be $5 \times \sigma^{\text{LD}}_{\text{res}}$, reaches $\SI{1}{\electronvolt}$ and therefore allows the resolution of energies smaller than the band gap of GaAs.
The monotonically decaying background with about $10^9$ events for a kg-year exposure used in the main text is here further suppressed by a factor $10^{-4}$ for better visualization of the peaks. 
The cross-section sensitivity including background is then derived by scanning the cross section for different DM masses with a confidence limit set to $90\%$.\\
The inclusion of a realistic background allows us to compare our BSE cross-section sensitivity in NaI to previous predictions by Zema et al.~\cite{Zema_light_detec:2024} based on QEdark~\cite{Essig_DM_e_semicond:2016}, where screening and excitonic effects were neglected and the band gap energy was used for the onset of both absorption and emission. 
For better comparison, here we choose the previously used value $\epsilon_l =10\%$ for the combined light yield and collection efficiency.
As shown in Fig.~\ref{reach_BSE_vs_QEdark_vs_DAMIC}, for the same background suppression of $\lambda_{\text{bck}}=10^{-4}$, the inclusion of electron-hole interactions improves the prediction of the cross-section sensitivity for NaI over the entire mass range such that the sensitivity reaches beyond the previous prediction where screening was neglected. This influence of the excitonic effects was already pointed out in the discussion of the reach without background in Fig.~\ref{fig_reach_only_theory}.
Fig.~\ref{reach_BSE_vs_QEdark_vs_DAMIC} also shows recent exclusion limits on DM-electron interactions of the DAMIC-M experiment~\cite{DAMIC-M_exclusion:2025}. We see that to probe a new parameter space with GaAs and NaI, further suppression of the background, improvement of light efficiency, and a larger exposure of the targets are required. 

\bibliography{references}